\documentclass[useAMS,usenatbib]{mn2e}
\usepackage{epsfig}
\newcommand{\beq}{\begin{equation}}
\newcommand{\eeq}{\end{equation}}

\newcommand{\apj}{ApJ}
\newcommand{\apjl}{ApJL}
\newcommand{\apjs}{ApJS}
\newcommand{\aj}{AJ}
\newcommand{\mnras}{MNRAS}

\newdimen\hssize
\newdimen\hdsize 
\title[Interactions, star formation and AGN activity]
{Interactions, star formation and AGN activity}

\author[Li et al.]
{Cheng Li$^{1,2}$\thanks{E-mail: leech@mpa-garching.mpg.de},
 Guinevere Kauffmann$^{2}$, Timothy M. Heckman$^{3}$,
 Simon D.M. White$^{2}$, \newauthor Y. P. Jing$^{1}$ \\
${^1}$MPA/SHAO Joint Center for Astrophysical Cosmology at
      Shanghai Astronomical Observatory,
      Nandan Road 80, Shanghai 200030, China \\
${^2}$Max Planck Institut f\"ur Astrophysik,
      Karl-Schwarzschild-Strasse 1, 85748 Garching, Germany \\
${^3}$ Department of Physics and Astronomy,
       Johns Hopkins University, Baltimore, MD 21218
}

\begin{document}

\date{Accepted ........ Received ........; in original form ........}

\pagerange{\pageref{firstpage}--\pageref{lastpage}} \pubyear{2007}

\maketitle

\label{firstpage}

\begin {abstract} It has long  been known that galaxy interactions are
associated  with enhanced  star formation.  In a  companion  paper, we
explored this connection  by applying a variety of  statistics to SDSS
data.  In particular, we showed  that specific star formation rates of
galaxies  are higher  if they  have close  neighbours.  Here  we apply
exactly the same  techniques to AGN in the  survey, showing that close
neighbours are not associated  with any similar enhancement of nuclear
activity. Star formation  is enhanced in AGN with  close neighbours in
exactly the same  way as in inactive galaxies,  but the accretion rate
onto  the black hole,  as estimated  from the  extinction-corrected [O
{\sc iii}] luminosity, is not influenced by the presence or absence of
companions. Previous  work has shown that galaxies  with more strongly
accreting black holes contain more young stars in their inner regions.
This  leads us  to conclude  that star  formation induced  by  a close
companion and star formation  associated with black hole accretion are
distinct  events.  These  events  may  be part  of  the same  physical
process, for  example a merger,  provided they are separated  in time.
In this  case, accretion onto the  black hole and  its associated star
formation  would occur only  after the  two interacting  galaxies have
merged.   The major caveat  in this  work is  our assumption  that the
extinction-corrected [O {\sc iii}] luminosity is a robust indicator of
the  bolometric luminosity  of the  central  black hole.   It is  thus
important to  check our  results using indicators  of AGN  activity at
other wavelengths.

\end {abstract}


\section {Introduction}

In \citet[][hereafter Paper I]{Li-07}  we used a variety of statistics
to study the  small-scale clustering properties of a  sample of $10^5$
star-forming galaxies  selected from the  data release 4 of  the Sloan
Digital  Sky  Survey  \citep[SDSS;][]{York-00}.   We  cross-correlated
star-forming  galaxies  with reference  samples  drawn  from the  main
spectroscopic  survey and  from  the photometric  catalogue (which  is
complete down to significantly  fainter limiting magnitudes).  We also
calculated  the  average  enhancement  in  star formation  rate  as  a
function of the projected  distance to companion galaxies. Our results
supported  a picture  in which  enhanced star  formation  activity and
galaxy interactions are closely  linked.  At the highest specific star
formation  rates, almost  all  galaxies  were found  to  have a  close
companion or be in the  process of merging.  Galaxies with the highest
specific star formation rates were also found to be more concentrated,
suggesting  that  these merger-driven  interactions  were building  up
their   bulge  components.    These   results  are   in  accord   with
gas-dynamical simulations of mergers of disk galaxies, which show that
substantial fraction  of the gas in  the galaxy can  be driven towards
the center of the galaxy by gravitational torques, where it can fuel a
central starburst \citep{Negroponte-White-83,Mihos-Hernquist-96}.

The discovery of a tight correlation between black hole mass and bulge
mass       or        velocity       dispersion       in       galaxies
\citep{Gebhardt-00,Ferrarese-Merritt-00}     has     motivated    many
theoretical  models  in  which  accretion  onto black  holes  and  AGN
activity are assumed  to be closely linked to  galaxy interactions and
mergers \citep[see  for example][]{Kauffmann-Haehnelt-00, Cattaneo-01,
Granato-01,  Wyithe-Loeb-02,   DiMatteo-03,  Cattaneo-05,  Hopkins-05,
Kang-05, Bower-06,  Croton-06}.  In many of  these models ,  a few per
cent of the gas in the  merging galaxies is accreted by the black hole
and  shines with a  radiative efficiency  of around  10\% over  a time
scale of  a few $\times  10^7$ years.  With these  simple assumptions,
many of the observed features of the cosmological evolution of quasars
appear to be reasonably  well reproduced.  It is therefore interesting
to explore whether we can find any empirical evidence for a connection
between  merging and  AGN  activity in  the  real Universe.   Previous
observational  studies have  yielded contradictory  results \citep[see
for   example][]   {Petrosian-82,   Dahari-84,   Dahari-85,   Keel-85,
Fuentes-Williams-Stocke-88,    Virani-00,    Schmitt-01,    Miller-03,
Grogin-05, Waskett-05, Koulouridis-06,  Serber-06}.  It is thus useful
to carry out  a careful analysis of a large  and homogeneous sample of
AGN.

In  \citet{Li-06},  we presented  results  on  the  clustering of  AGN
selected from the SDSS. Our study  did not yield any evidence that AGN
activity is triggered by  interactions with nearby companions. Instead
we found that AGN are clustered {\em more weakly} than control samples
of non-AGN  on scales between 100  kpc and 1 Mpc.  We interpreted this
anti-bias as evidence  that AGN activity is favoured  in galaxies that
are  located  at the  centres  of their  own  dark  matter halos,  and
disfavoured in  ``satellite galaxies''  that orbit together  with many
other galaxies within a common halo.

The  purpose  of  this  short   paper  is  to  bring  our  results  on
star-forming  galaxies  and  AGN  together.  We  analyze  star-forming
galaxies  and AGN  in exactly  the  same way,  using the  same set  of
statistics, and we compare and  contrast the results that are obtained
for the two  kinds of galaxy. We then discuss  the implications of our
results for  understanding the connection between  star formation, AGN
activity and galaxy interactions

\begin{figure*}   
\centerline{
\psfig{figure=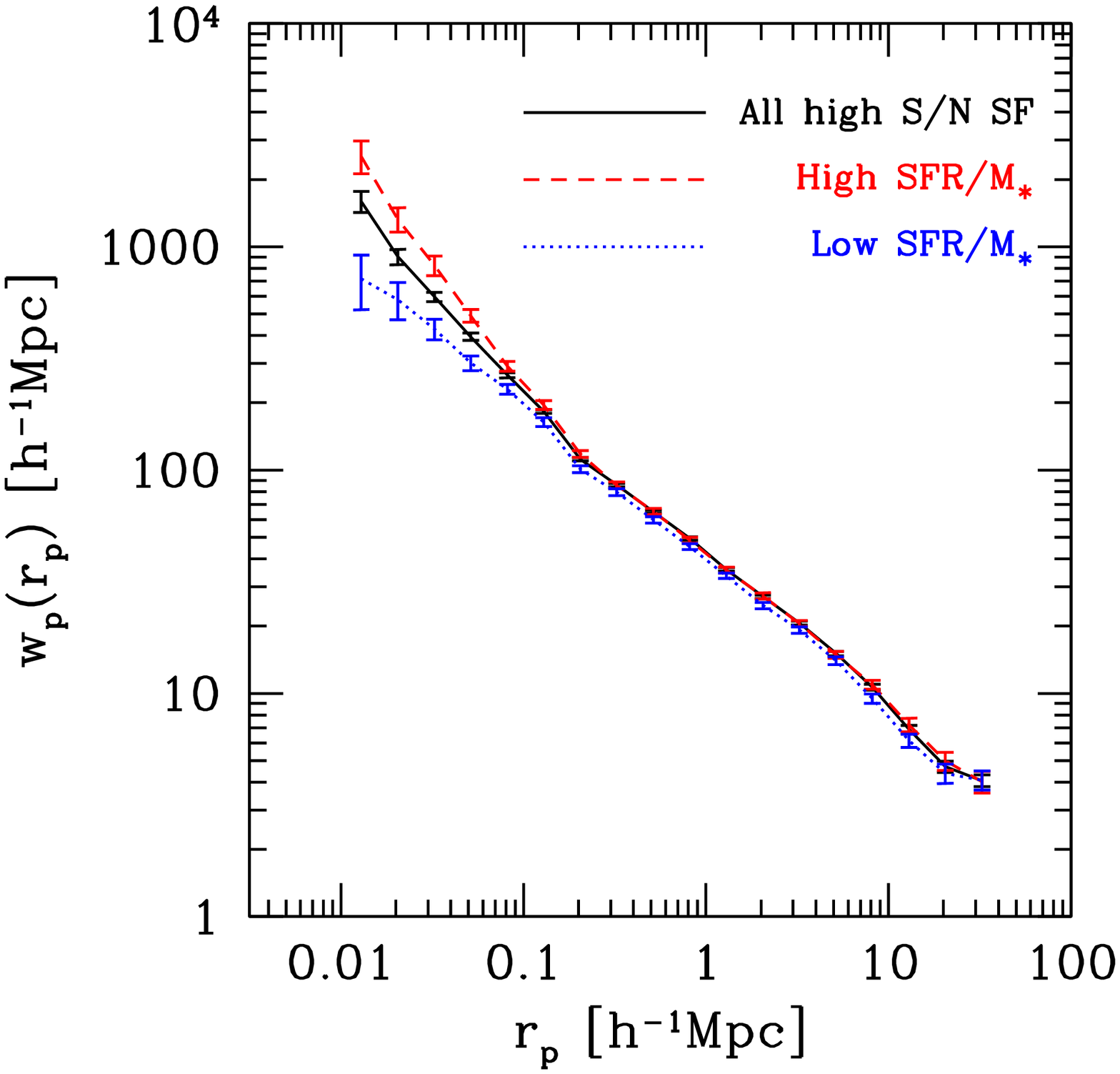,clip=true,width=0.5\textwidth}
\psfig{figure=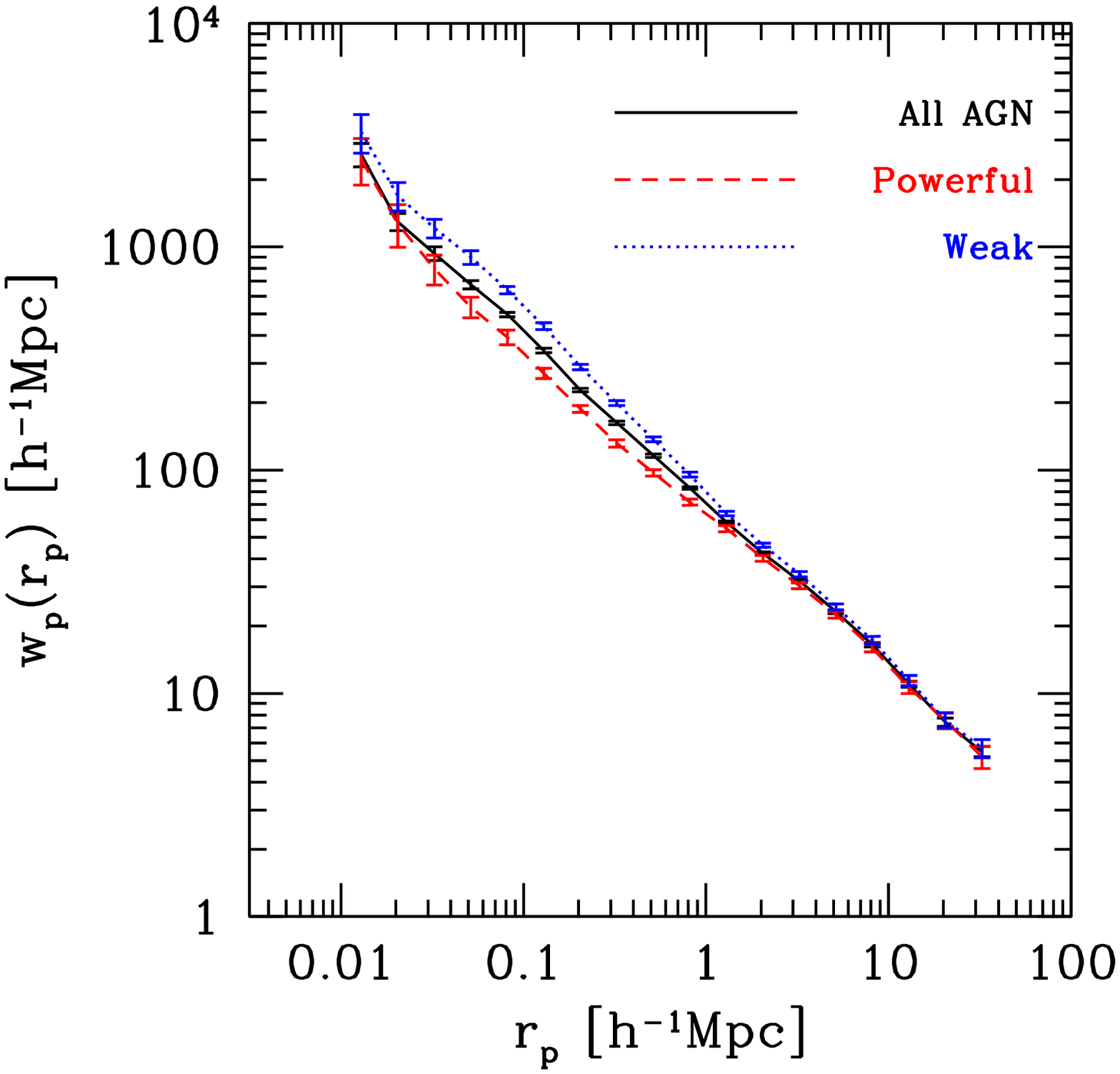,clip=true,width=0.5\textwidth}
}
\caption{Projected    redshift-space    two-point    cross-correlation
functions $w_p(r_p)$  between all  high S/N star-forming  galaxies and
our  reference sample  (left) and  between all  AGN and  our reference
sample (right).   Different lines correspond  to star-forming galaxies
with different specific star formation rates, or to AGN with different
accretion  rates, as  indicated.  See  the text  for  a more  detailed
description.}
\label{fig:wrp}
\end{figure*}

\section {Indicators of Star Formation and AGN Activity}

Our basic indicator of star formation activity in galaxies is the {\em
specific star formation rate} , which is defined as the star formation
rate of the galaxy measured within the 3 arcsecond SDSS fibre aperture
divided by  the stellar mass of  the galaxy measured  within this same
aperture.   Extinction-corrected star  formation  rates are  estimated
using    a   variety    of    emission   lines    as   described    in
\citet{Brinchmann-04}.

Our indicator of  AGN activity is the quantity  L[O {\sc iii}]/M$_{\rm
BH}$, where L[O {\sc iii}]  is the [O {\sc iii}]$\lambda5007$ emission
line luminosity in solar units and $M_{BH}$ is the black hole mass (in
solar masses) estimated from  the measured stellar velocity dispersion
of   the    galaxy   using    the   fitting   formula    provided   by
\citet{Tremaine-02}.      As    discussed     in    \citet[][hereafter
K03]{Kauffmann-03}, the  contribution to the [O  {\sc iii}] luminosity
from [H II] regions is small in nearby AGN.  The [O {\sc iii}] line is
expected  to  come almost  exclusively  from  the narrow-line  region,
making  it   an  excellent  probe   of  accretion  onto   the  central
supermassive  black hole.  Additional  support for  the [O  {\sc iii}]
line  luminosity as  an accretion  rate indicator  is provided  by the
study of a hard X-ray selected sample of 47 AGN by \citet{Heckman-05},
which showed  that the  3-20 keV and  [O {\sc iii}]  luminosities were
well-correlated over a range of about 4 orders of magnitude.

The main  drawback of the  [O {\sc iii}]  emission line is that  it is
affected by dust extinction.   \citet{Hopkins-05} developed a model in
which the  lifetime and visibility of quasars  is intimately connected
with  obscuration  inside merging  systems.  In  their model,  mergers
funnel gas  to galaxy centres,  fuelling starbursts and  feeding black
hole growth, but the quasars remain ``buried'' until accretion-related
feedback  disperses  the obscuring  material  and  they are  revealed.
Eventually,  activity  ceases as  the  remnants  settle  down and  the
accretion  rate  drops.  In  such  a model  obscuration  effects  vary
dramatically as a system ages. Here, we attempt to correct the [O {\sc
iii}] luminosity  for dust obscuration  using the value of  the Balmer
decrement measured  from the  galaxy spectrum (see  K03).  This  is of
course  a crude correction,  because the  H$\alpha$ and  H$\beta$ line
line fluxes are  likely to be dominated by emission  from H II regions
that may  be spread over  a larger spatial  scale than the  gas clouds
that  contribute to  high-ionization emission  lines such  as  [O {\sc
iii}].  Nevertheless, we believe that  correcting for dust in this way
will improve  the correlation between  our AGN activity  indicator and
the bolometric luminosity of the central black hole.

We  note  that  the majority  of  AGN  in  our sample  are  relatively
low-luminosity objects with [O {\sc iii}] line luminosities well below
those  of typical  high-redshift  quasars.  \citet{Heckman-04}  showed
that high-mass  black holes are  accreting very little at  the present
day;  only black  holes  of  $10^8$ $M_{\odot}$  or  less continue  to
accrete  and to  grow at  a significant  rate.  The  SDSS  optical AGN
sample includes 4680 objects with  $\log$ L[O {\sc iii}]/M$_{BH} > 1$.
This  cut selects  the $\sim  5\%$  of our  AGN sample  with the  most
rapidly growing black holes.   According to the calibration between [O
{\sc   iii}]   line   luminosity   and   accretion   rate   given   in
\citet{Heckman-04},   and  taking   account  the   average  extinction
correction that  we apply here  to the [O  {\sc iii}] line,  the black
holes  in these  galaxies are  accreting at  approximately a  tenth of
Eddington or more.  Even though these AGN may have significantly lower
[O  {\sc iii}] luminosities  than high-redshift  quasars, they  are as
powerful  as these quasars  when their  luminosities are  expressed in
Eddington  units.  We  thus  believe that  they  can provide  valuable
insight into the processes  that are responsible for triggering strong
AGN activity.

\begin{figure*}
\centerline{\psfig{figure=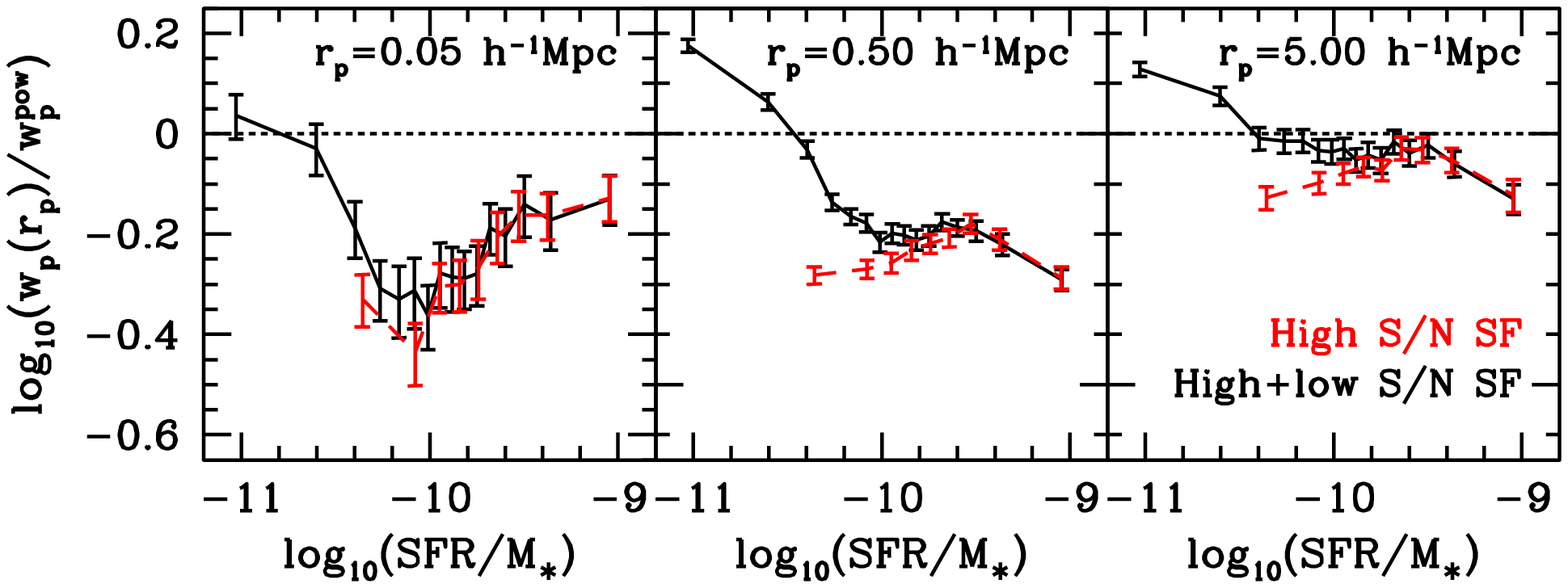,clip=true,width=\textwidth}}
\centerline{\psfig{figure=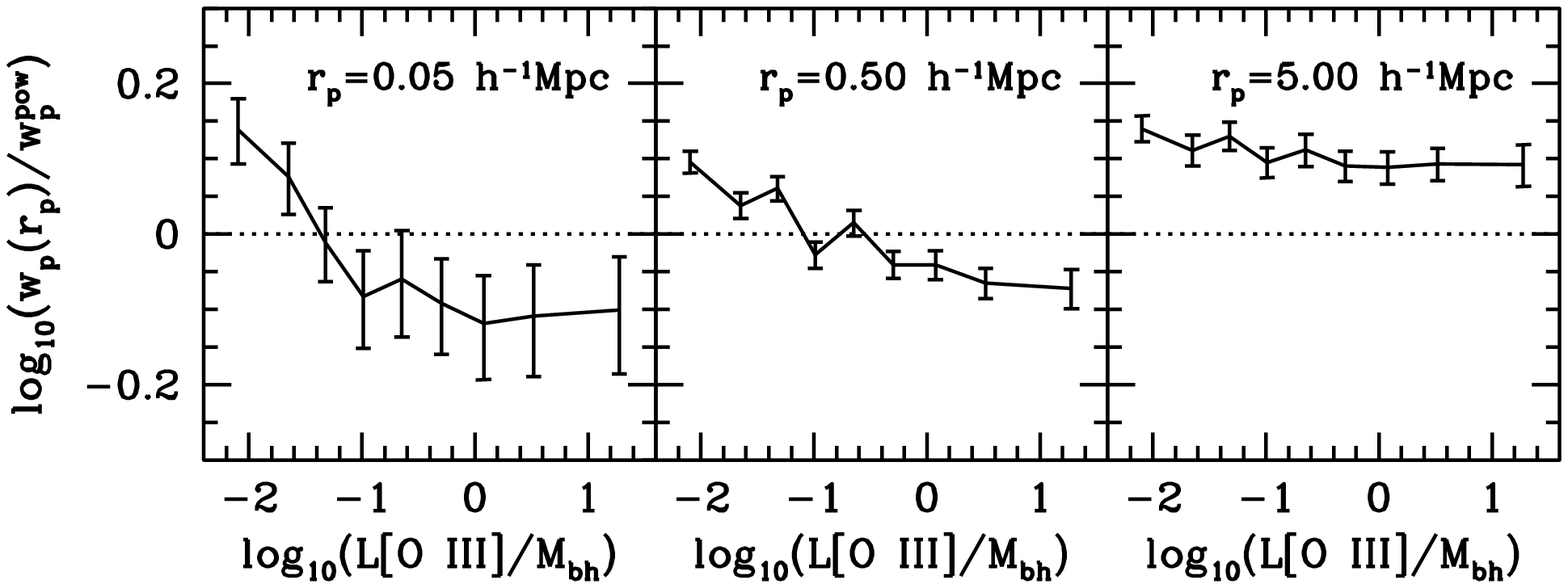,clip=true,width=\textwidth}}
\caption{Projected two-point cross-correlations $w_p(r_p)$, normalized
by  the   expectation  for   a  power-law  3-D   correlation  function
$\xi(r)=(r/5h^{-1}Mpc)^{-1.8}$  and  measured   at  the  $r_p$  values
indicated   in   each   panel,   are   plotted  as   a   function   of
$\log_{10}SFR/M_\ast$ for star-forming galaxies  (top panels) and as a
function  of  $\log_{10}(L$[O  {\sc  iii}]$/M_{BH})$ for  AGN  (bottom
panels).}
\label{fig:wrp_scales}
\end{figure*}

\begin{figure*}
\centerline{
\psfig{figure=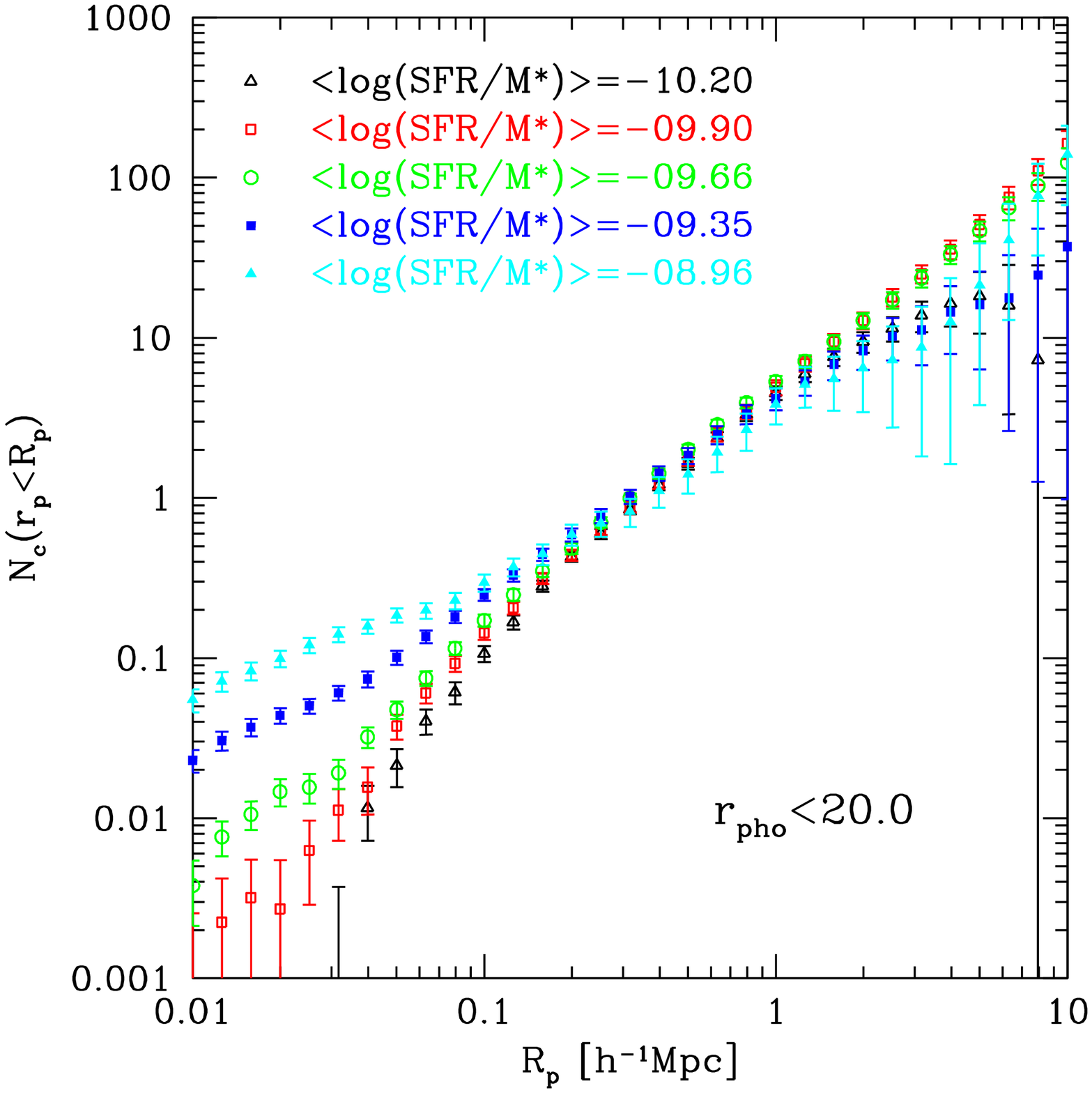,clip=true,width=0.5\textwidth}
\psfig{figure=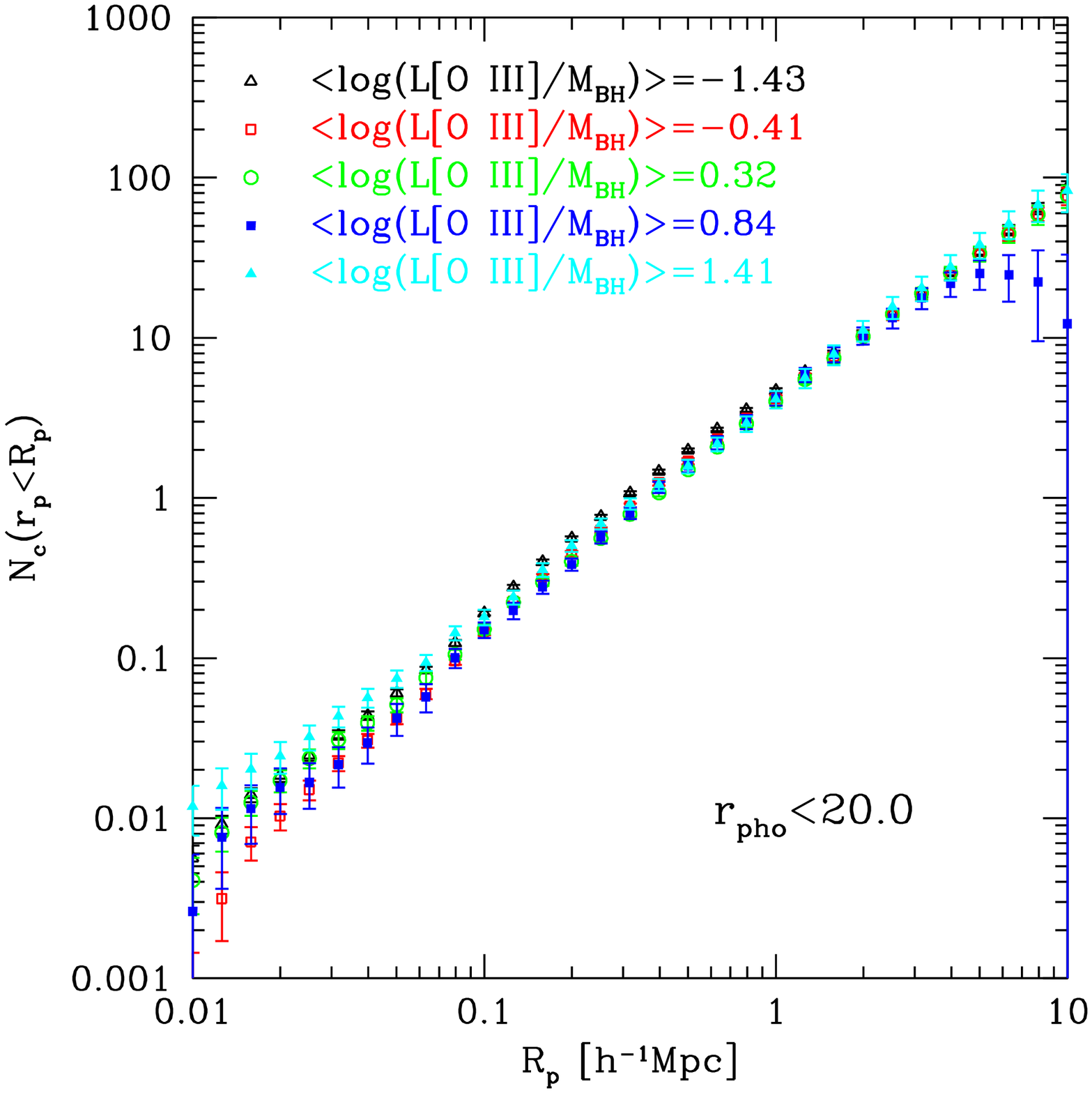,clip=true,width=0.5\textwidth} }
\caption{Average counts  of galaxies in  the photometric sample  to an
$r$-band limiting  magnitude of $r_{lim}=20$ within  a given projected
radius $r_p$  from the star-forming  galaxies (left) and from  the AGN
(right). Different symbols are  for star-forming galaxies in different
intervals of  $\log_{10}SFR/M_\ast$ or for AGN  in different intervals
of $\log_{10}(L$[O {\sc iii}]/M$_{BH})$, as indicated.}
\label{fig:counts}
\end{figure*}

\section {Results}

Our sample of star-forming galaxies  is the same as described in Paper
I. The  sample of AGN used here  is the same as  in \citet{Li-06}.  In
short, the  parent sample  is composed of  397,344 objects  which have
been spectroscopically  confirmed as  galaxies and have  data publicly
available in the SDSS Data Release~4 \citep{Adelman-McCarthy-06}.  AGN
and  star-forming galaxies are  selected from  the subset  of galaxies
with $S/N > 3$ in  the four emission lines [O {\sc iii}]$\lambda$5007,
H$\beta$,  [N II]$\lambda$6583 and  H$\alpha$, following  the criteria
proposed by K03.

Our  methodology for  evaluating  the clustering  properties of  these
galaxies  has already  been  described in  Paper  I, so  we will  turn
directly to our results.

\begin{figure*}
\centerline{
\psfig{figure=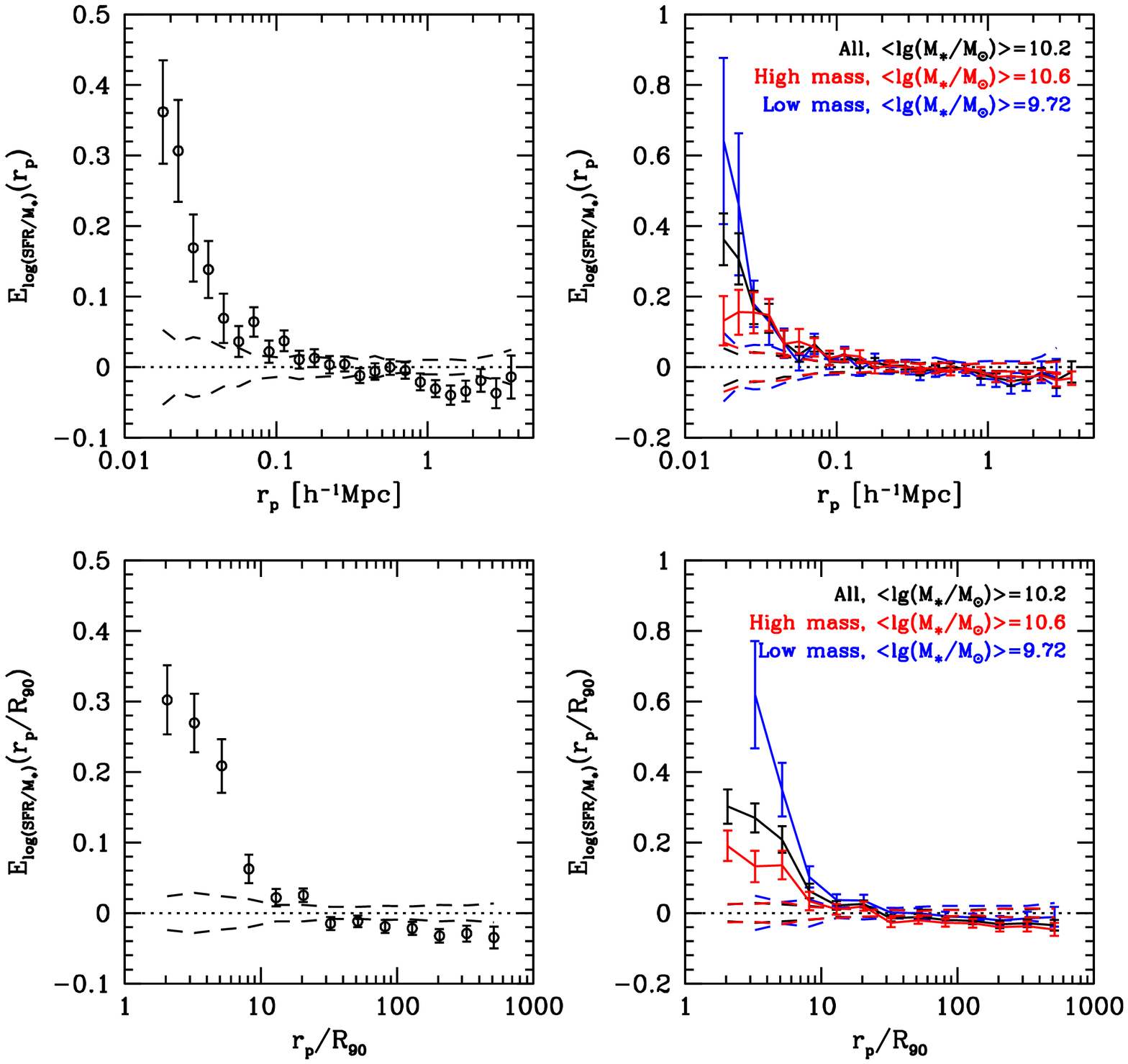,clip=true,width=0.5\textwidth}
\psfig{figure=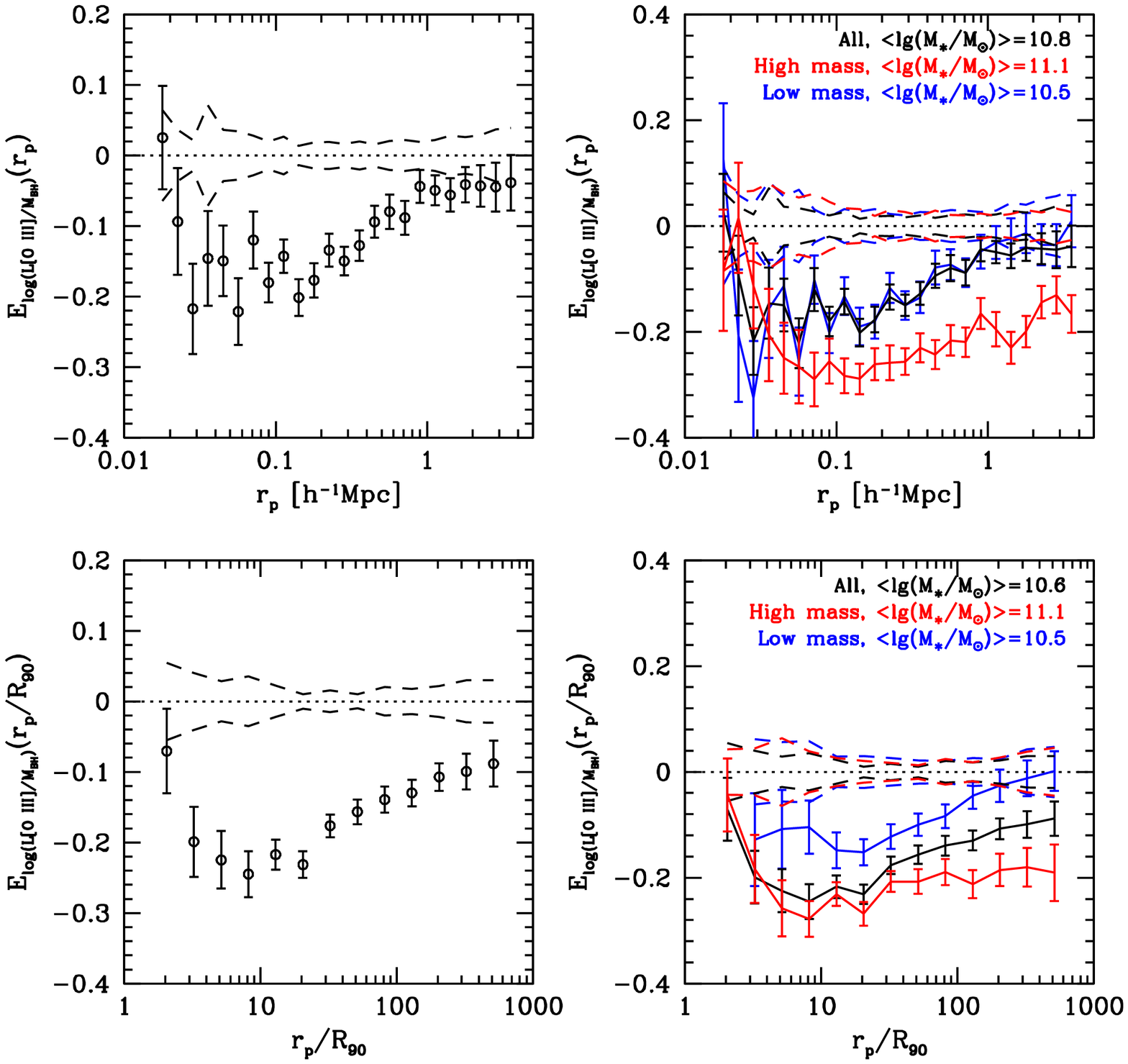,clip=true,width=0.5\textwidth}
}
\caption{Enhancement    in   $\log_{10}SFR/M_\ast$   for    high   S/N
star-forming galaxies (left) and in $\log_{10}L$[O {\sc iii}]$/M_{BH}$
for AGN (right), as a function of the projected separation $r_p$ (top)
and as a function of the scaled separation $r_p/R_{90}$ (bottom). Here
$r  <17.6$  for  the  central  star-forming galaxies  and  AGN,  while
$r<19.0$  for the reference  sample of  galaxies from  the photometric
catalogue.  }
\label{fig:naef}
\end{figure*}

\begin{figure*}
\centerline{
\psfig{figure=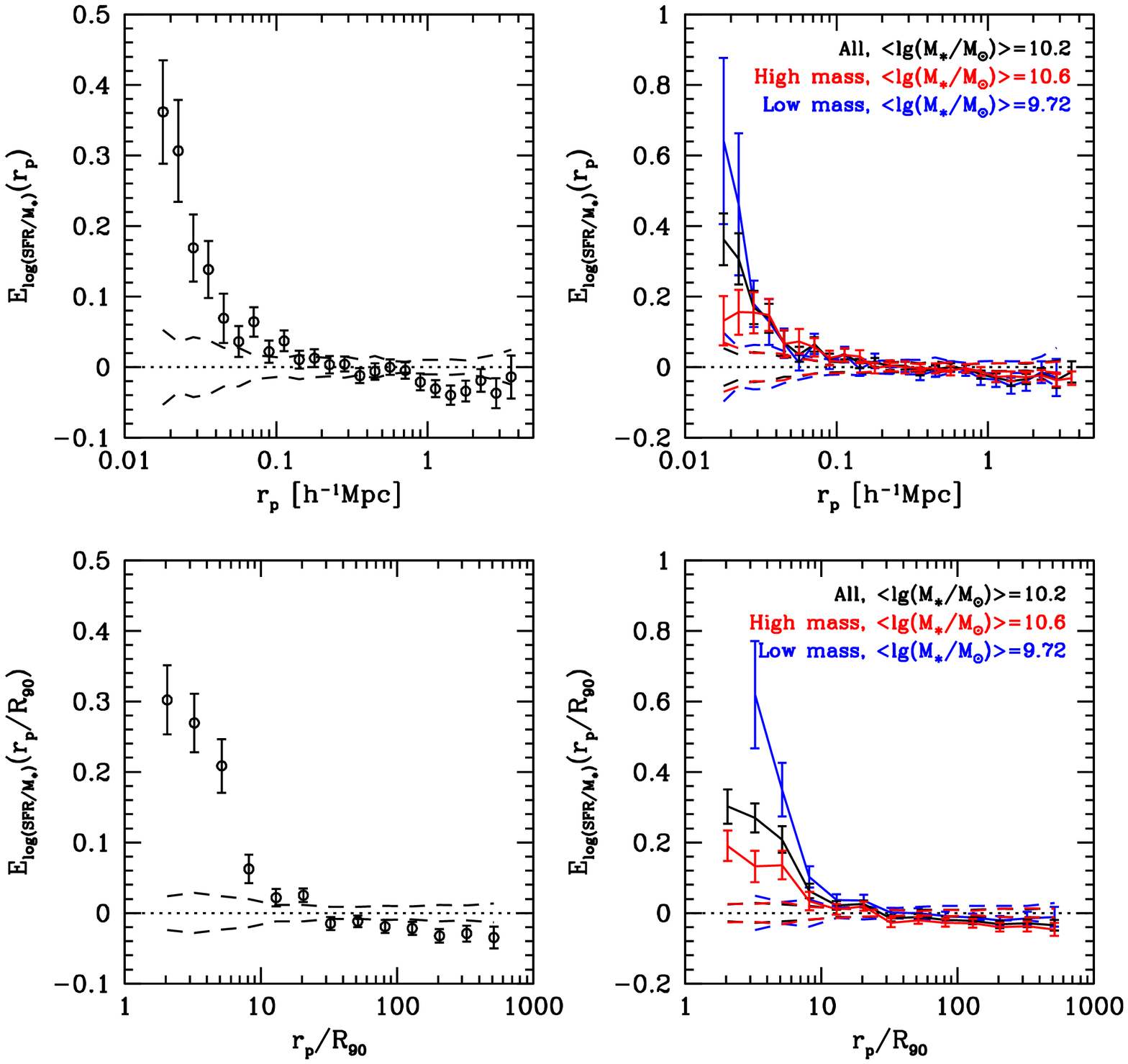,clip=true,width=0.5\textwidth}
\psfig{figure=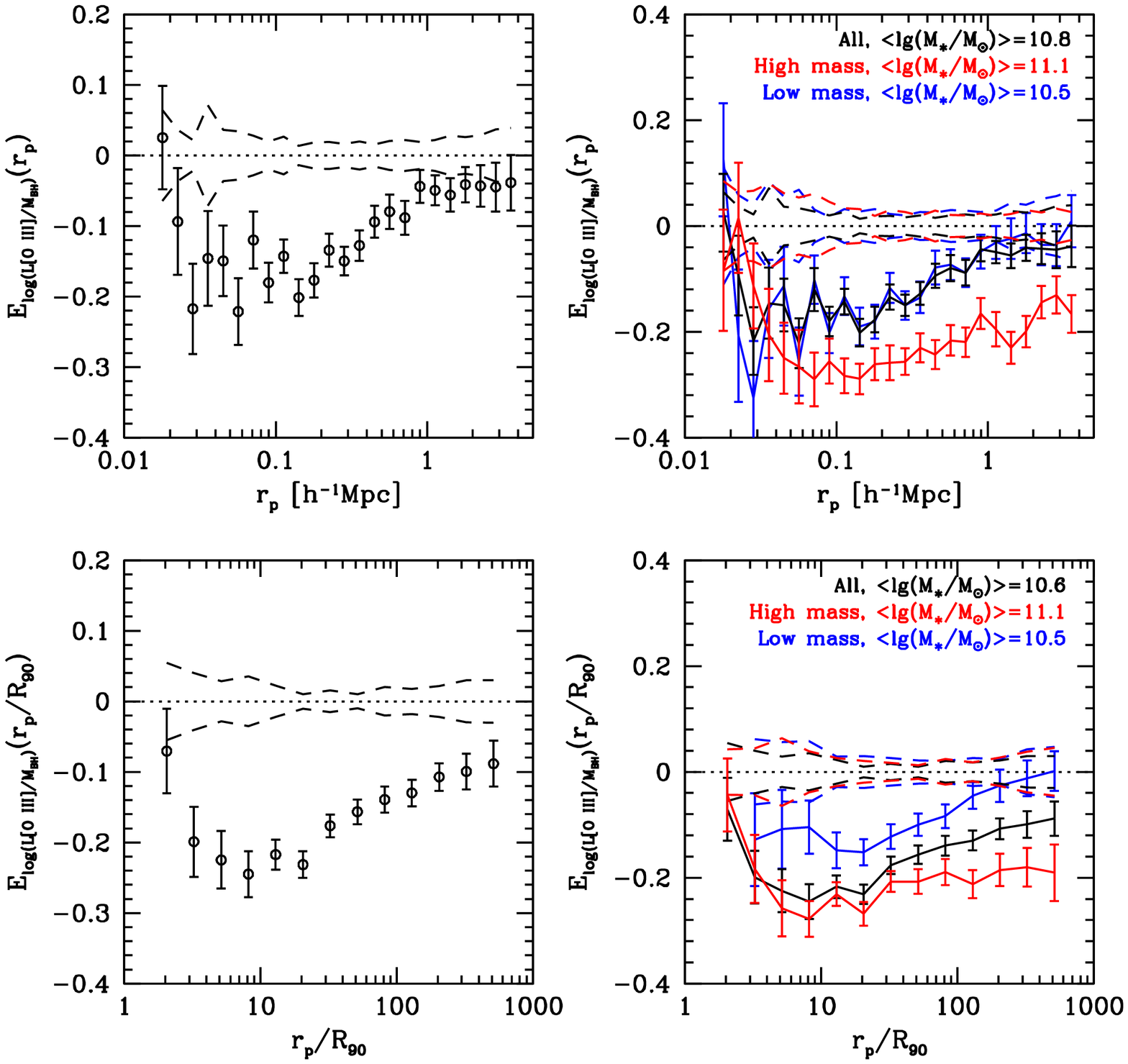,clip=true,width=0.5\textwidth}
}
\caption{Enhancement    in   $\log_{10}SFR/M_\ast$   for    high   S/N
star-forming galaxies (left) and in $\log_{10}L$[O {\sc iii}]$/M_{BH}$
for AGN (right), as a  function of the scaled separation $r_p/R_{90}$,
for objects in the different stellar mass ranges as indicated.  }
\label{fig:naef_mass}
\end{figure*}

\begin{figure*}
\centerline{\psfig{figure=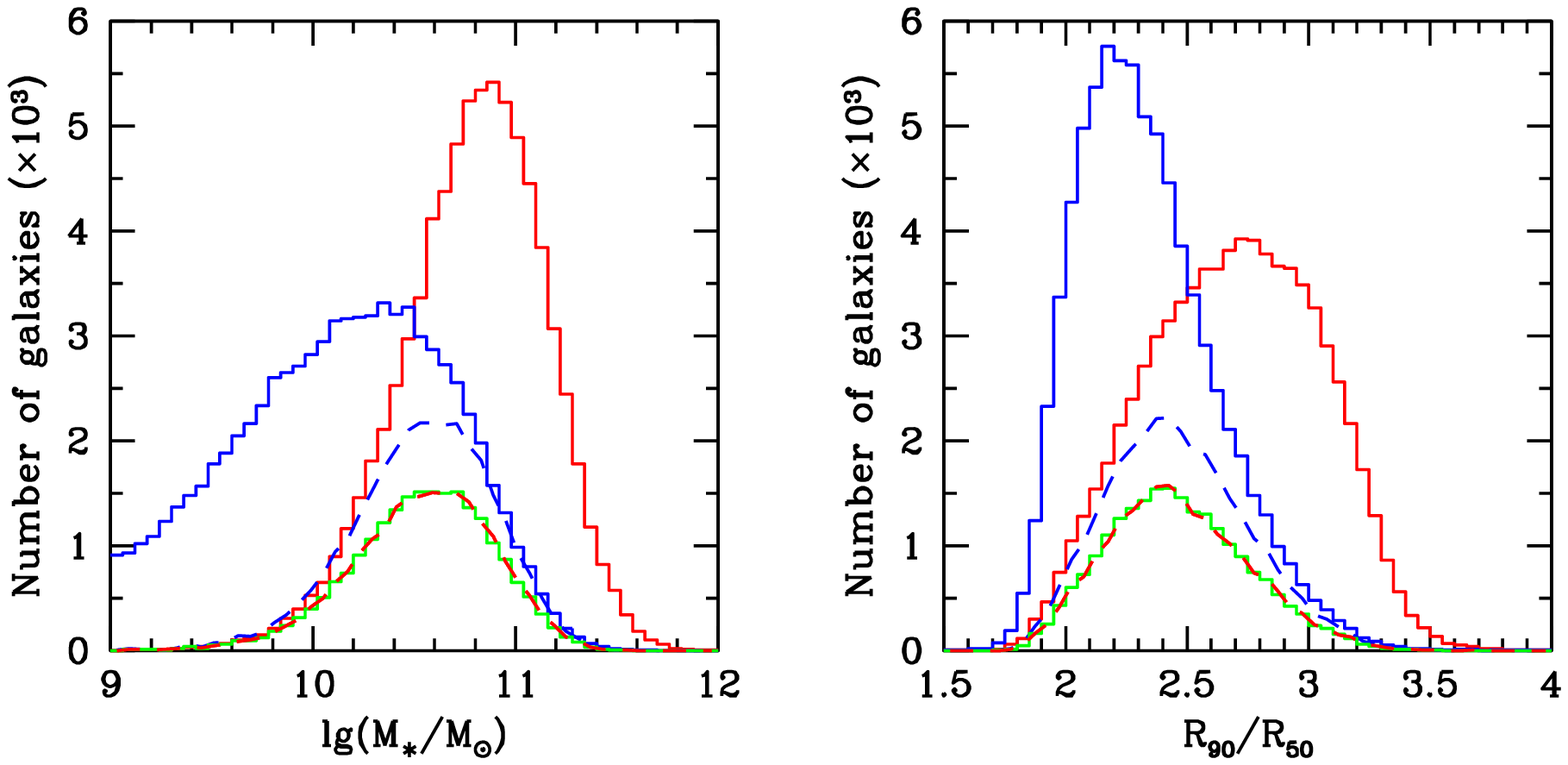,clip=true,width=\textwidth}}
\caption{Distribution of  stellar mass (left)  and concentration index
(right) for the  full star-forming sample (blue solid  line), the full
AGN sample (red solid line), the star-forming sample matched to AGN in
redshift, stellar  mass and concentration (blue dashed  line), and the
AGN  matched to  star-forming  galaxies in  the  same properties  (red
dashed  line).  The   green  solid  line  is  also   for  the  matched
star-forming  sample, but  is normalized  as  to have  the same  total
number of objects as the matched AGN sample.}
\label{fig:mass_con_his}
\end{figure*}

\subsection {Cross-correlation functions}

In Figure~\ref{fig:wrp},  we compare projected  redshift-space 2-point
cross-correlation  functions  (2PCF)  $w_p(r_p)$  for  our  sample  of
star-forming galaxies  (left) and AGN (right).  Results  for the whole
sample are  plotted in black.  Results  for the 25\%  of galaxies with
the  smallest values  of  SFR/M$_*$ and  L[O  {\sc iii}]/M$_{BH}$  are
plotted in  blue.  Results for the  25\% of galaxies  with the highest
values of  SFR/M$_*$ and L[O  {\sc iii}]/M$_{BH}$ are plotted  in red.
The left panel  shows that galaxies with high  specific star formation
rates are more strongly clustered than galaxies with low specific star
formation  rates on scales  less than  100 $h^{-1}$  kpc and  that the
difference in  clustering amplitude increases  as one goes  to smaller
values  of  $r_p$.  The  right  panel  shows  that AGN  display  quite
different  behaviour.  AGN with  the highest  values of  the Eddington
parameter L[O  {\sc iii}]/M$_{BH}$ are {\em more  weakly} clustered on
scales between 30 kpc and 1 Mpc.  At the smallest values of $r_p$, the
clustering amplitude exhibits no dependence on AGN power.

These    results    are    shown    again   in    more    detail    in
Figure~\ref{fig:wrp_scales}, where  we plot  the amplitude of  the the
2PCF  as  a function  of  SFR/$M_*$  and as  a  function  of L[O  {\sc
iii}]/M$_{BH}$ for three different  values of the projected separation
($r_p =  0.05$, $0.5$ and $5  h^{-1}$ Mpc).  We extend  the results to
lower values of  SFR/$M_*$ using the sample of  low $S/N$ star-forming
galaxies described in  Paper I. It is interesting  that the trends are
qualitatively  similar for  the two  kinds  of object,  except on  the
smallest scales where the clustering amplitude increases substantially
for the large  values of SFR/$M_*$, but exhibits  no dependence on the
Eddington parameter  L[O {\sc iii}]/M$_{BH}$.  We note  that this lack
of dependence extends up to the very highest values of $\log$ L[O {\sc
iii}]/M$_{BH}$,  corresponding to  black holes  that are  accreting at
more than a tenth of the Eddington rate.

For  both star-forming  galaxies and  AGN, the  small-scale clustering
amplitude increases  strongly towards low values of  SFR/M$_*$ and L[O
{\sc iii}]/M$_{BH}$.   This is  most likely {\em  not} a  signature of
mergers  or interactions.   In a  recent paper,  \citet{Barton-07} use
cosmological N-body simulations to show that a substantial fraction of
close galaxy pairs reside in  cluster or group-size halos.  It is well
known that  the star  formation rates of  galaxies are lower  in these
environments \citep[e.g.][]{Balogh-99}, and  the same is probably true
of the average accretion rate  onto the central black hole (K03).  The
upturn  in  clustering  amplitude  for  galaxies with  low  values  of
SFR/$M_*$ and for weak AGN  is thus a signature of galaxies ``shutting
down'' in the most massive halos.

\begin{figure*}                                                      
\centerline{\psfig{figure=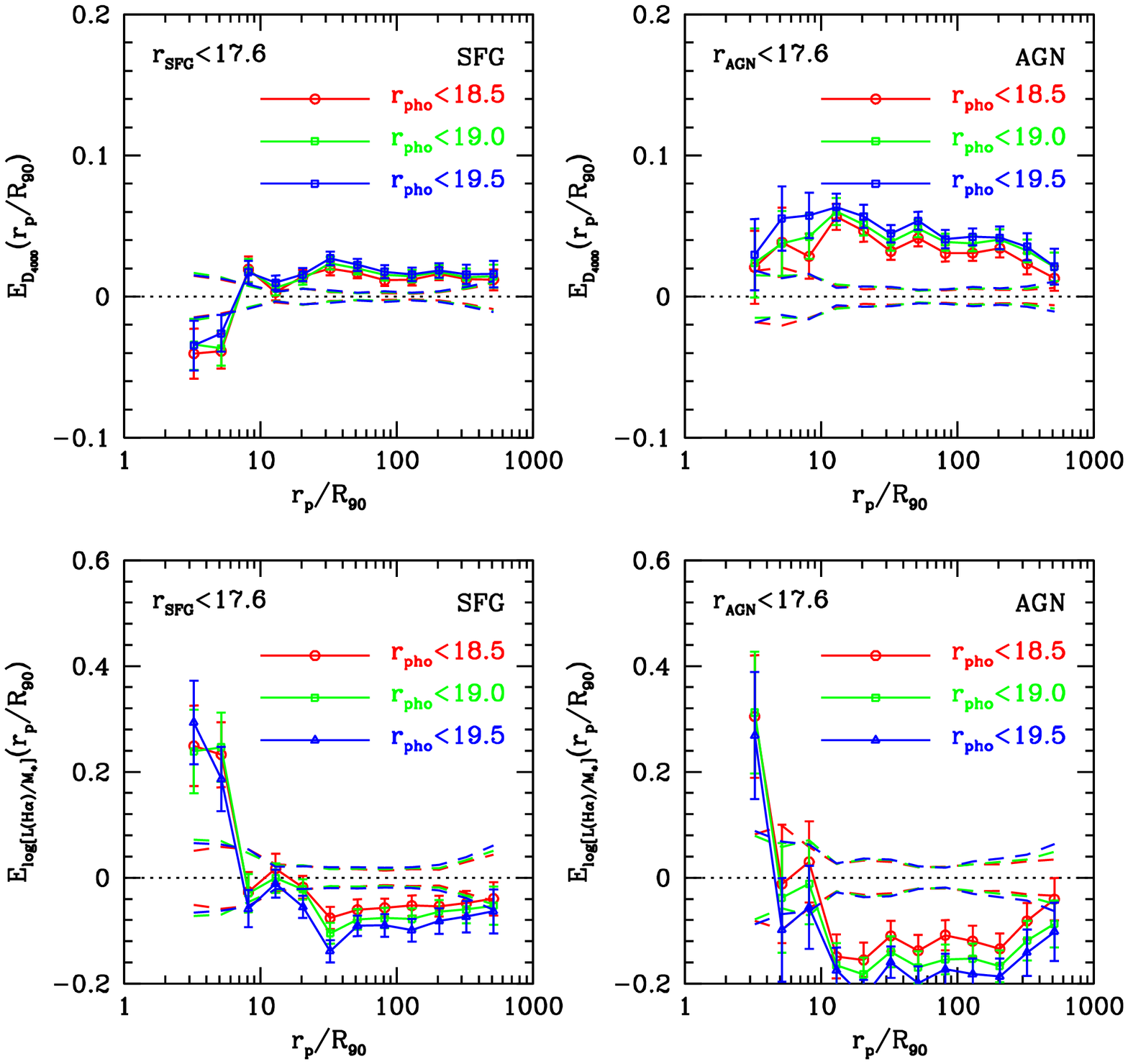,clip=true,width=\textwidth}}
\caption{Enhancement in $\log_{10}L(H_\alpha)/M_\ast$ as a function of
the scaled separation $r_p/R_{90}$, for high S/N star-forming galaxies
(left) and for  AGN (right).  The AGN and  star-forming galaxy samples
are   matched  with  each   other  in   redshift,  stellar   mass  and
concentration index.  The limiting  magnitude of AGN  and star-forming
galaxies is fixed at $r=17.6$, but the apparent magnitude limit of the
reference galaxy sample is varied as indicated.}
\label{fig:enhancement_d4k_lumhalpha}
\end{figure*}

\subsection {Neighbour Counts}

As in  Paper I, we compute  the number of galaxies  in the photometric
reference sample in the vicinity  of star-forming galaxies and AGN and
we make a statistical correction for the effect of chance projections.
We have  trimmed the  AGN and star-forming  samples so that  they each
have  the  same  distribution   in  redshift  and  stellar  mass.   In
Figure~\ref{fig:counts}  we  plot  the  average  background-subtracted
neighbour count  within a given  value of the projected  radius $r_p$.
Results are shown for  star-forming galaxies in different intervals of
SFR/$M_*$  and  AGN in  different  intervals  of $\log_{10}(L$[O  {\sc
iii}]/M$_{BH})$.  The  photometric reference sample  is always limited
at  $r_{pho}$= 20.0.  We see  a  clear trend  for an  increase in  the
average  number   of  close   neighbours  around  the   most  strongly
star-forming galaxies, but not around the most powerful AGN.

\subsection {"Enhancement" functions}

In     Paper    I,     we    computed     the    background-corrected,
neighbour-count-weighted enhancement  in specific star  formation rate
for galaxies as a function of projected neighbour distance, The reader
is referred to  Paper I for definitions and  for further details about
the method.

Figure~\ref{fig:naef}  compares  the   enhancement  in  specific  star
formation rate  (left) with  the enhancement in  the level  of nuclear
activity,  as measured  by the  quantity log  L[O  {\sc iii}]/M$_{BH}$
(right). The top panels show  the enhancement functions as function of
projected separation  $r_p$, while in the bottom  panel the separation
is scaled  by dividing by  R90, the radius  that encloses 90\%  of the
$r$-band light of the galaxy.

As can be seen, AGN  behave very differently to star forming galaxies.
Accretion onto  the black hole  is {\em suppressed} for  galaxies with
companions with projected separations between $\sim 100$ kpc -- 1 Mpc,
and there is  no evidence that the nuclear  activity level is enhanced
above  the mean at  small neighbour  separations.  The  most plausible
explanation  for the suppression  on intermediate  scales is  that the
majority  of AGN  with  close  neighbours are  located  in groups  and
clusters  and the suppression  of nuclear  activity is  a larger-scale
environmental effect, similar to the morphology-density or SFR-density
relations.  This hypothesis is in accord with the AGN clustering model
of  \citet{Li-06}, in which  AGN activity  is suppressed  in satellite
galaxies relative to central galaxies.

In Figure~\ref{fig:naef_mass},  we show how  the enhancement functions
depend  on  stellar  mass  for  both  the  star-forming  and  the  AGN
samples. The  suppression of AGN activity in  galaxies with companions
with  $r_p=0.1-1$ Mpc does  depend on  stellar mass  -- the  effect is
clearly much stronger for the most massive galaxies. However, there is
again no evidence that AGN activity  is triggered by the presence of a
close companion  in the  way seen for  star-formation activity  at all
stellar masses.  We  find similar trends when we  split our AGN sample
by a structural  parameter such as the concentration  index $C$. There
is  stronger  intermediate  scale  suppression for  more  concentrated
galaxies,  but  no  evidence   for  significant  triggering  by  close
companions in any of our subsamples.

Our goal in this paper is  to understand whether there is a connection
between  interactions,  enhanced  star  formation,  and  enhanced  AGN
activity for  SDSS galaxies.  In Paper  I, we argued  that there  is a
clear  connection  between   galaxy  interactions  and  enhanced  star
formation.   We also  argued  that galaxy  interactions  are not  only
sufficient  but   also  {\em  necessary}  to   trigger  the  strongest
starbursts.   In   K03,  we   demonstrated  that  powerful   AGN  have
significantly  younger stellar  populations than  non-AGN of  the same
stellar mass.  Clearly, there is a connection between interactions and
enhanced star formation, and there  is a connection between strong AGN
activity and  enhanced star formation.   The third link,  a connection
between interactions and strong AGN activity, appears to be missing..

To illustrate this more clearly, we have created subsamples of AGN and
high $S/N$ star-forming galaxies that are closely matched in redshift,
stellar  mass and  concentration  index $C$.  This  is illustrated  in
detail  in Figure~\ref{fig:mass_con_his}.  As  can be  seen, the  full
star-forming   and   AGN  samples   differ   substantially  in   their
distributions of  both stellar mass  and concentration index.  We have
shown that  the enhancement functions are quite  strongly dependent on
these  two  parameters.  Thus, if  we  want  to  make a  {\em  direct}
comparison between  star-forming galaxies and  AGN, we need  to create
matched  subsamples.  We  can then  investigate whether  these matched
subsamples show  identical trends in  star formation enhancement  as a
function   of  neighbour   distance.    Our  results   are  shown   in
Figure~\ref{fig:enhancement_d4k_lumhalpha}.   Because the luminosities
of the  higher-ionization emission lines are strongly  affected by the
central  source,  star  formation   rates  cannot  be  estimated  very
accurately for AGN.  We focus instead on the H$\alpha$ luminosity of a
galaxy normalized  to its  stellar mass $M_*$.   As discussed  in K03,
H$\alpha$ line luminosities are very  similar in AGN and in samples of
star-forming galaxies  that are matched  in 4000 \AA\  break strength,
proving that most  of the H$\alpha$ emission arises  from H II regions
in both cases.

\begin{figure}
\centerline{\psfig{figure=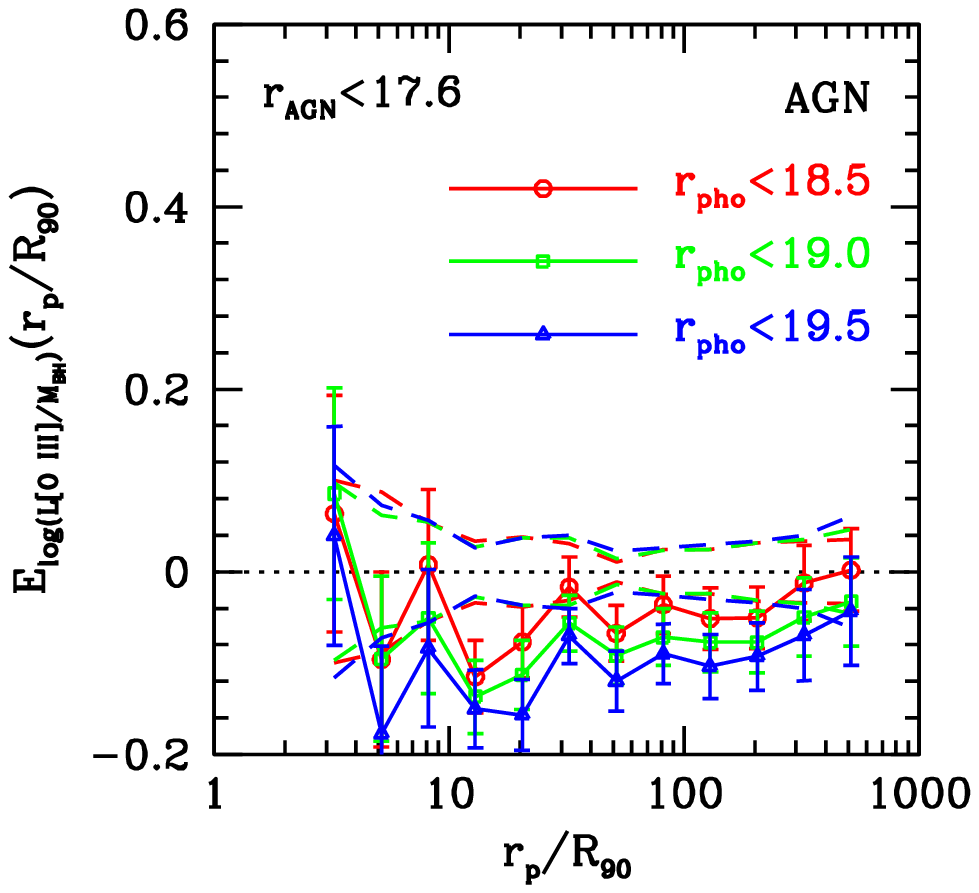,clip=true,width=0.5\textwidth}}
\caption{Enhancement  in  $\log_{10}L$[O   {\sc  iii}]$/M_{BH}$  as  a
function  of the  scaled  separation $r_p/R_{90}$,  for  AGN that  are
closely matched  to the high $S/N$ star-forming  galaxies in redshift,
stellar  mass and  4000  \AA\ break  index  D$_{4000}$.  The  limiting
magnitude  of the  AGN  sample  is fixed  at  $r_{AGN}=17.6$, but  the
apparent  magnitude limit  for  the reference  galaxies  is varied  as
indicated.  }
\label{fig:naef_mass_con_matched}
\end{figure}

Figure~\ref{fig:enhancement_d4k_lumhalpha} shows that on small scales,
the behaviour of this indicator  as a function of scaled separation is
the same in  the star-forming galaxies and in  the AGN.  This suggests
that interactions are regulating star formation rates in our sample of
AGN in exactly the same way as in our sample of star-forming galaxies!
In Figure~\ref{fig:naef_mass_con_matched}, we  plot the enhancement in
L[O  {\sc  iii}]/M$_{BH}$  for   the  AGN  contained  in  the  matched
sample.  The suppression  in  nuclear activity  level at  intermediate
scales is no longer quite as strong as in our full sample, because the
AGN in the matched sample  have significantly lower stellar masses and
concentrations. However, we still do not see any significant trend for
nuclear activity to increase as  the separation between the galaxy and
its neighbours  gets smaller.  We conclude that,  unlike enhanced star
formation, accretion onto the black  hole is not directly connected to
interactions for local AGN.

\section {Conclusions}

The basic conclusion of this paper is very simple. A strong connection
between galaxy  interactions and enhanced  star formation is  found no
matter  what  statistic  we  employ.  However, we  fail  to  find  any
corresponding relation between enhanced AGN activity and interactions.
The  interpretation  of  what   this  implies  for  the  starburst-AGN
connection is far from obvious.

We  know  from previous  work  \citep[e.g.][; K03]{Kauffmann-03}  that
galaxies with  powerful AGN tend to  have younger-than-average stellar
populations.   K03 found that  AGN of  all luminosities  reside almost
exclusively  in  massive galaxies  and  have  distributions of  sizes,
stellar surface mass densities  and concentrations that are similar to
those of  ordinary galaxies of  the same mass.   The hosts of  the AGN
with high  [O {\sc  iii}] luminosities, on  the other hand,  have much
younger mean stellar  ages than ``normal'' galaxies of  the same mass.
This establishes that there is a physical connection between accretion
onto the  central black hole  and the presence  of young stars  in the
inner galaxy.

There is also a  clear physical connection between galaxy interactions
and  enhanced levels  of star  formation. This  is true  not  only for
``normal'' galaxies, but also for the  host galaxies of the AGN in our
samples.  Figure 7 shows that star formation rates in AGN hosts with a
close  companion are enhanced  in exactly  the same  way as  in normal
star-forming  galaxies.  However,  we  consistently fail  to find  any
evidence  that the  black hole  accretion rate  , as  measured  by the
quantity $L$[O {\sc iii}]$/M_{BH}$, increases  if a galaxy has a close
companion.

These results  are summarized in Figure  9.  This plot  shows how star
formation in  AGN is  enhanced in the  plane of accretion  rate ($\log
L$[O {\sc iii}]/M$_{BH}$) versus  projected separation between the AGN
and  its neighbours.   Results  are  shown for  AGN  that are  closely
matched to high $S/N$  star-forming galaxies in redshift, stellar mass
and concentration  index.  Each bin  is colour-coded according  to the
average  value of $\log$  L(H$\alpha$)/$M_\ast$ for  all the  AGN that
fall into the given range  in log L[O {\sc iii}]/$M_{BH}$, weighted by
the true number  of companions at the projected  distance $r_p$.  This
weighted average value of log L(H$\alpha$)/$M_*$ is then scaled by the
mean value for  the sample as a whole.  In the  upper bar, the scaled,
weighted  average  value  of  log(H$\alpha$)/$M_*$) is  plotted  as  a
function of projected  separation for the same sample  of AGN. This is
thus the  average of  the 2-D  plot over all  accretion rates  at each
companion distance.

The  enhancement  in  star  formation  as a  function  of  $L$[O  {\sc
iii}]$/M_{BH}$ is clearly seen in this diagram.  It is also clear from
Figure 9  that the average value  of $L$[O {\sc  iii}]$/M_{BH}$ in our
sample has no dependence on neighbour distance. The apparent narrowing
in the distribution  at low values of the  scaled separation is simply
the result of poorer sample statistics.  Finally , it is striking that
the  degree to  which  star formation  is  enhanced as  a function  of
projected   separation  between   the  AGN   and  its   companions  is
considerably weaker than the enhancement that occurs as the black hole
accretion  rate increases.  This point  can be  seen by  comparing the
upper bar in Figure 9 with the main panel.

These  results   lead  us  to  the  conclusion   that  star  formation
enhancement due  to a close  companion and star  formation enhancement
due  to an accreting  black hole  are {\it  two separate  and distinct
events.} These two events may  be part of the same underlying physical
process (such as a merger),  provided they are well separated in time.
In this  case, accretion onto the  black hole and  its associated star
formation  would occur only  after the  two interacting  galaxies have
already merged.

In  a recent  paper, Yuan  et  al (2007,in  preparation) have  studied
starburst  and AGN activity  in infrared-selected  galaxies, including
ULIRGs.  These authors  study  how the  relative  fraction of  objects
classified as starburst  or AGN change as the  merger event progresses
.  They  find that  there  is no  significant  change  in fraction  of
galaxies with AGN activity until  the two galactic nuclei have merged.
The single  nucleus post-merger phase  is divided into  three classes,
according  to  the  K-band  core-to-total ratio  and  morphology.  The
authors find that for ULIRGS there is a particular stage (the "diffuse
merger" stage) where the nuclei have merged but not yet formed a core,
where there is a substantial rise in composite HII-AGN activity.  Once
the core forms (the  so-called "compact merger" stage), this composite
activity changes into mostly pure Seyfert activity.  In another recent
paper, \citet{Ellison-08}  have analyzed the star  formation rates and
AGN activity for a sample of  1716 paired galaxies selected  from  the
SDSS,  and  also  find  that  if  AGN  activity  is  associated   with 
merger-triggered  star formation, the  timescales  must  be longer for 
the former than for the latter.

These  results appear  the  support  the hypothesis  that  there is  a
significant time  delay between the  onset of the interaction  and the
phase  in  which  the black  hole  is  able  to accrete.   This  delay
hypothesis  can  be  tested  by  using empirical  measures  of  galaxy
morphology that are more sensitive  to the later stages of the merging
event.         Examples        include        asymmetry        indices
\citep{Abraham-06,Conselice-Bershady-Jangren-00}         or        the
``lopsidedness'' parameter \citep{Reichard-07}.

A major  caveat in  all the  analysis presented in  this paper  is the
assumption that the extinction-corrected [O {\sc iii}] luminosity is a
reasonably  robust  indicator  of  the bolometric  luminosity  of  the
central black hole.   This is the only indicator  of accretion that is
available  to us  for our  sample of  AGN from  the Sloan  Digital Sky
Survey.  It will be extremely important to check the results presented
in this paper using indicators of AGN activity at other wavelengths.

\section*{Acknowledgements} 

CL is  supported by the Joint Postdoctoral  Programme in Astrophysical
Cosmology  of  Max  Planck  Institute for  Astrophysics  and  Shanghai
Astronomical Observatory.  CL and YPJ are supported by NSFC (10533030,
10643005,  10633020),  by  the  Knowledge Innovation  Program  of  CAS
(No. KJCX2-YW-T05), and by  973 Program (No.2007CB815402).  CL, GK and
SW would like  to thank the hospitality and  stimulating atmosphere of
the Aspen Center  for Physics while this work  was being completed and
Roderik Overzier for useful discussions.

Funding for  the SDSS and SDSS-II  has been provided by  the Alfred P.
Sloan Foundation, the Participating Institutions, the National Science
Foundation, the  U.S.  Department of Energy,  the National Aeronautics
and Space Administration, the  Japanese Monbukagakusho, the Max Planck
Society, and  the Higher Education  Funding Council for  England.  The
SDSS Web  Site is  http://www.sdss.org/.  The SDSS  is managed  by the
Astrophysical    Research    Consortium    for    the    Participating
Institutions. The  Participating Institutions are  the American Museum
of  Natural History,  Astrophysical Institute  Potsdam,  University of
Basel,   Cambridge  University,   Case  Western   Reserve  University,
University of Chicago, Drexel  University, Fermilab, the Institute for
Advanced   Study,  the  Japan   Participation  Group,   Johns  Hopkins
University, the  Joint Institute  for Nuclear Astrophysics,  the Kavli
Institute  for   Particle  Astrophysics  and   Cosmology,  the  Korean
Scientist Group, the Chinese  Academy of Sciences (LAMOST), Los Alamos
National  Laboratory, the  Max-Planck-Institute for  Astronomy (MPIA),
the  Max-Planck-Institute  for Astrophysics  (MPA),  New Mexico  State
University,   Ohio  State   University,   University  of   Pittsburgh,
University  of  Portsmouth, Princeton  University,  the United  States
Naval Observatory, and the University of Washington.

\begin{figure*}
\centerline{
\psfig{figure=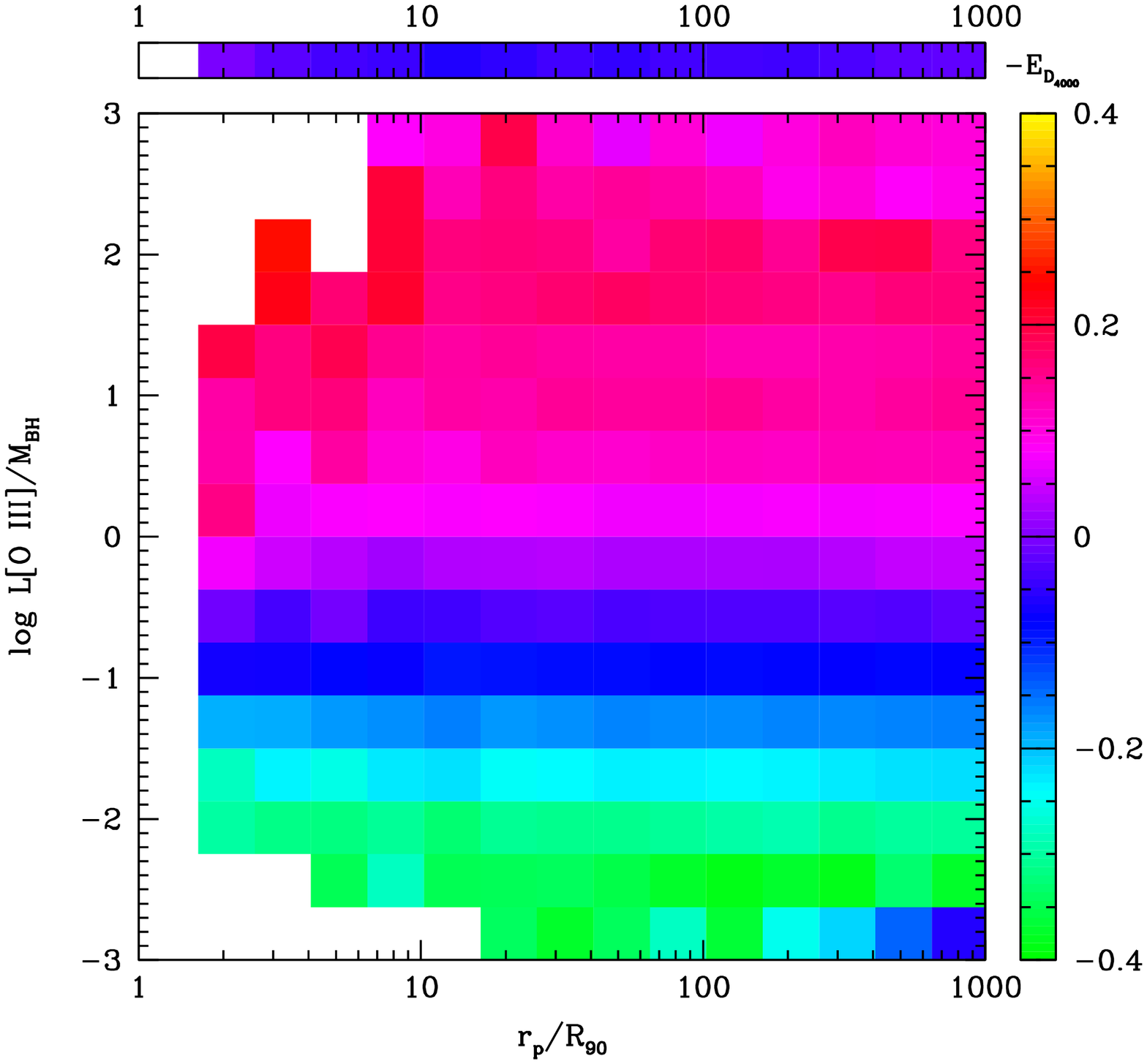,clip=true,width=0.5\textwidth}
\psfig{figure=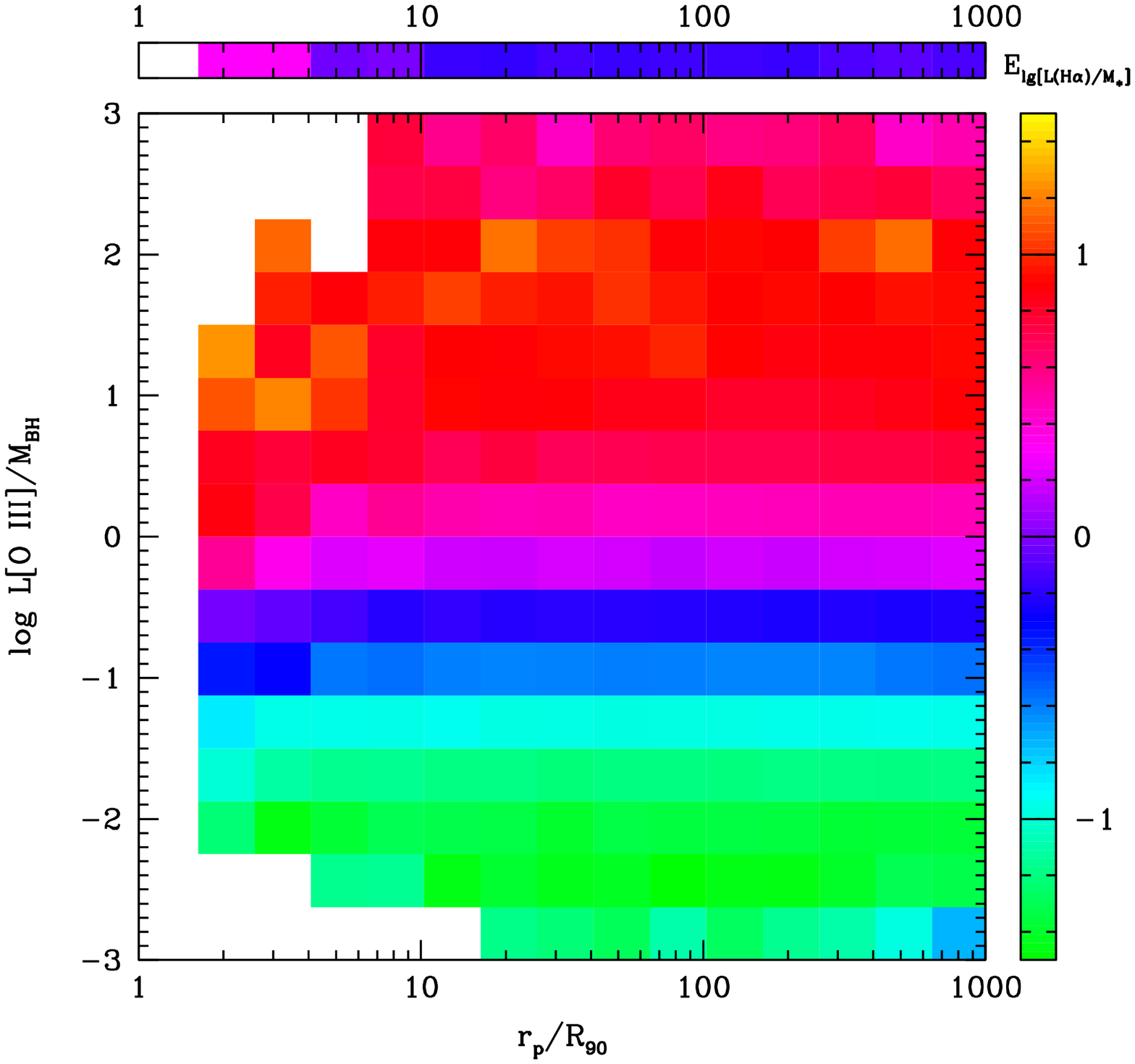,clip=true,width=0.5\textwidth}
}
\caption{ The enhancement in star  formation for AGN is plotted in the
plane  of  accretion  rate  ($L$[O  {\sc  iii}]/M$_{BH}$)  versus  the
projected separation between an AGN host and its neighbours (scaled to
the 90\%  light radius of the  host).  Results are shown  for AGN that
are closely  matched to high $S/N$ star-forming  galaxies in redshift,
stellar  mass  and  concentration  index.  Each  bin  is  colour-coded
according to the average  value of $-$D$_{4000}$ (the left-hand panel)
or $\log  L$(H$\alpha$)/$M_*$ (the right-hand  panel) for all  the AGN
that  fall into  the given  range in  $\log L$[O  {\sc iii}]/$M_{BH}$,
weighted by the true number of companions at projected distance $r_p$.
This    weighted   average   value    of   $-$D$_{4000}$    or   $\log
L$(H$\alpha$)/$M_\ast$ is then scaled by the mean value for the sample
as a whole.  In each panel,  the colour coding is indicated by the bar
on the  right. The bar at  the top shows the  scaled, weighted average
value of  $-$D$_{4000}$ or $\log L$(H$\alpha$)/$M_\ast$  as a function
of projected separation for the same sample of AGN. This is simply the
vertical average of the data in the main panel.}
\label{fig:binned}
\end{figure*}


\bsp
\label{lastpage}

\end{document}